\documentclass[aps,twocolumn,prd,showpacs,showkeys,preprintnumbers,superscriptaddress,bibnotes,floatfix,longbibliography]{revtex4-2}

\usepackage{graphicx}
\usepackage{bm}
\usepackage{times}
\usepackage{hyperref}
\usepackage{slashed}
\usepackage{color}
\usepackage{aas_macros}
\usepackage{nicefrac}
\usepackage{soul}

\usepackage{slashed}
\usepackage{lipsum}
\usepackage{subfigure}
\usepackage{multirow}
\usepackage{amsmath}
\usepackage{array}  
\usepackage{varwidth} 
\usepackage{comment}

\hypersetup{
    pdfnewwindow=true,    
    colorlinks=true,      
    linkcolor=blue,       
    citecolor=blue,       
    filecolor=blue,      
    urlcolor=blue         
}
\graphicspath{{figures/}}
\newcommand{\orcid}[1]
{\begingroup
  \hypersetup{hidelinks}\href{https://orcid.org/#1}{\includegraphics[width=9pt]{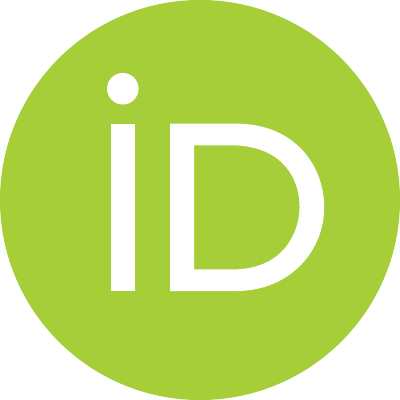}
} \endgroup}

\begin{document}

\title{Where is the End of the Cosmic-Ray Electron Spectrum?}

\author{Takahiro Sudoh \orcid{0000-0002-6884-1733}}
\email{takahiro\_sudoh@icloud.com}
\affiliation{Center for Cosmology and AstroParticle Physics (CCAPP), Ohio State University, Columbus, OH 43210, USA}
\affiliation{Department of Physics, Ohio State University, Columbus, OH 43210, USA}
\affiliation{Department of Astronomy, Ohio State University, Columbus, OH 43210, USA}
\affiliation{Graduate School of Artificial Intelligence and Science, Rikkyo University, Nishi-Ikebukuro 3-34-1, Toshima-ku, Tokyo 171-8501, Japan}

\author{John F. Beacom \orcid{0000-0002-0005-2631}}
\email{beacom.7@osu.edu}
\affiliation{Center for Cosmology and AstroParticle Physics (CCAPP), Ohio State University, Columbus, OH 43210, USA}
\affiliation{Department of Physics, Ohio State University, Columbus, OH 43210, USA}
\affiliation{Department of Astronomy, Ohio State University, Columbus, OH 43210, USA}

\date{\today}

\begin{abstract}
Detecting the end of the cosmic-ray (CR) electron spectrum would provide important new insights.  While we know that Milky Way sources can accelerate electrons up to at least $\sim$1~PeV, the observed CR electron spectrum at Earth extends only up to 5~TeV (possibly 20~TeV), a large discrepancy.  The question of the end of the CR electron spectrum has received relatively little attention, despite its importance.  We take a comprehensive approach, showing that there are multiple steps at which the observed CR electron spectrum could be cut off.  At the highest energies, the accelerators may not have sufficient luminosity, or the sources may not allow sufficient escape, or propagation to Earth may not be sufficiently effective, or present detectors may not have sufficient sensitivity.  For each step, we calculate a rough range of possibilities.  \textit{Although all of the inputs are uncertain, a clear vista of exciting opportunities emerges.}  We outline strategies for progress based on CR electron observations and auxiliary multi-messenger observations.  In addition to advancing our understanding of CRs in the Milky Way, progress will also sharpen sensitivity to dark matter annihilation or decay.
\end{abstract}

\maketitle


\section{Introduction}

While the Milky Way is known to host powerful accelerators of cosmic-ray (CR) hadrons (protons and nuclei), the nature of these sources is uncertain because CR directions are smeared by propagation through magnetic fields~\cite{1964ocr..book.....G, 1990acr..book.....B, 1990cup..book.....G, 2016crpp.book.....G}.  Even so, we gain important clues about possible sources from other observables, including the CR spectrum and composition.  One of the most important clues is the spectral break at $\sim$3~PeV (the ``knee") in the CR spectrum~\cite{2008ApJ...678.1165A, 2013APh....47...54A, 2019PhRvD.100h2002A, 2020APh...12002441L}, which is commonly interpreted as the maximum proton energy that can be reached in typical accelerators.  This sets a scale that should be predicted by theory; the fact that this energy is so high greatly restricts the properties of possible sources~\cite{2013A&ARv..21...70B, 2019IJMPD..2830022G, 2021Univ....7..324C}.

Similarly, while the Milky Way is also known to host powerful accelerators of CR leptons (electrons and positrons), the nature of these sources is also uncertain~\cite{1964ocr..book.....G, 1990acr..book.....B, 1990cup..book.....G, 2016crpp.book.....G}.  (Hereafter, we use electrons to mean the sum of electrons and positrons unless otherwise specified.)  CR electrons have a much lower flux and a steeper spectrum than CR hadrons~\cite{2010ApJ...722L..58S, 2008PhRvL.101z1104A, 2019PhRvL.122j1101A, 2019PhRvL.122d1102A, 2017PhRvD..95h2007A, 2018PhRvL.120z1102A, 2017Natur.552...63D, 2018PhRvD..98f2004A}, which is partially a consequence of CR electrons having greater energy losses~\cite{2011hea..book.....L}.  While this makes CR electron observations more challenging, it means that the possible sources are closer, which will allow more detailed observations.  Many aspects of CR electrons have received attention in the theoretical literature, e.g., see Refs.~\cite{1995PhRvD..52.3265A, 2004ApJ...601..340K, 2010ApJ...710..958K, 
2009JCAP...01..025H, 2009PhRvL.103e1101Y, 
2009arXiv0912.0264K, 2009APh....32..140G, 
2010PhRvD..82b3009C, 
2011ApJ...743L...7H, 2011ApJ...729...93K, 2012PhRvL.108z1101G, 2012arXiv1210.8180K, 
2012CEJPh..10....1P, 
2013ApJ...772...18L, 2013PhRvL.111u1101B, 
2014PhRvD..89h3007G, 
2016JCAP...12..025A, 2017ApJ...836..172F, 2017JCAP...09..029J, 2017PhRvD..96b3015Z, 2018PhRvD..97l3011C, 2018JCAP...11..045M, 2018PhRvD..98h3009H, 2018ApJ...866..143H, 2018PhRvL.121y1106L, 2018MNRAS.478.5660F, 
2018PhRvD..97l3005J, 
2018PhRvD..98f3008C, 2019PhRvD..99j3022R, 2019ApJ...876L...8B, 2019JCAP...04..024M, 
2019PhRvD..99d3005L, 2020JCAP...02..009F, 2021arXiv210208456N, 2021PhRvD.103h3010E, 2021PhRvD.104j3013F, 2021PhRvD.103k5010D, 2021PhRvD.104h3012D, 2021JCAP...02..030L, 2021PhRvD.104l3029E,
2021MNRAS.508.6142M, 2022PhRvD.105b3015C, 
2022ApJ...926....5A, 2023PhRvD.107j3021J,
2023PhRvD.107f3003C}.

In this paper, we address a central question about CR electrons --- where does their spectrum end? --- that has received less attention, likely because a clear end has not yet been observed.  In direct CR observations, electrons have been robustly detected up to 5~TeV and possibly 20~TeV~\cite{2019PhRvL.122j1101A, 2019PhRvL.122d1102A, 2017PhRvD..95h2007A, 2018PhRvL.120z1102A, 2017Natur.552...63D, 2018PhRvD..98f2004A}.  However, we have indirect indications that electrons are accelerated to much higher energies, based on observations of the emission they produce due to synchrotron losses and inverse-Compton scattering~\cite{2021Sci...373..425L, 2017Sci...358..911A, 2021ApJ...911L..27A, 2022ApJ...930..148B, 2023ApJ...945...66P, 2023ApJ...945...33P, 2023arXiv230607347W, 2018A&A...612A...2H}.  For example, luminous sources like the Crab accelerate electrons up to at least 1 PeV~\cite{2021Sci...373..425L} and less luminous (but more common) sources like Geminga accelerate electrons up to at least 100~TeV~\cite{2017Sci...358..911A}.  However, it is not known if such electrons significantly contribute to the observed CR electron spectrum.  Determining the end of the CR electron spectrum would provide important new insights about the sources and propagation of all CR particles.  Near the end of the spectrum, it may be that only one source contributes, which would allow astronomy even without directionality.  And beyond the end, the sensitivity to exotic sources like dark matter would greatly improve.

\begin{figure*}
    \centering
    \includegraphics[width=2\columnwidth]{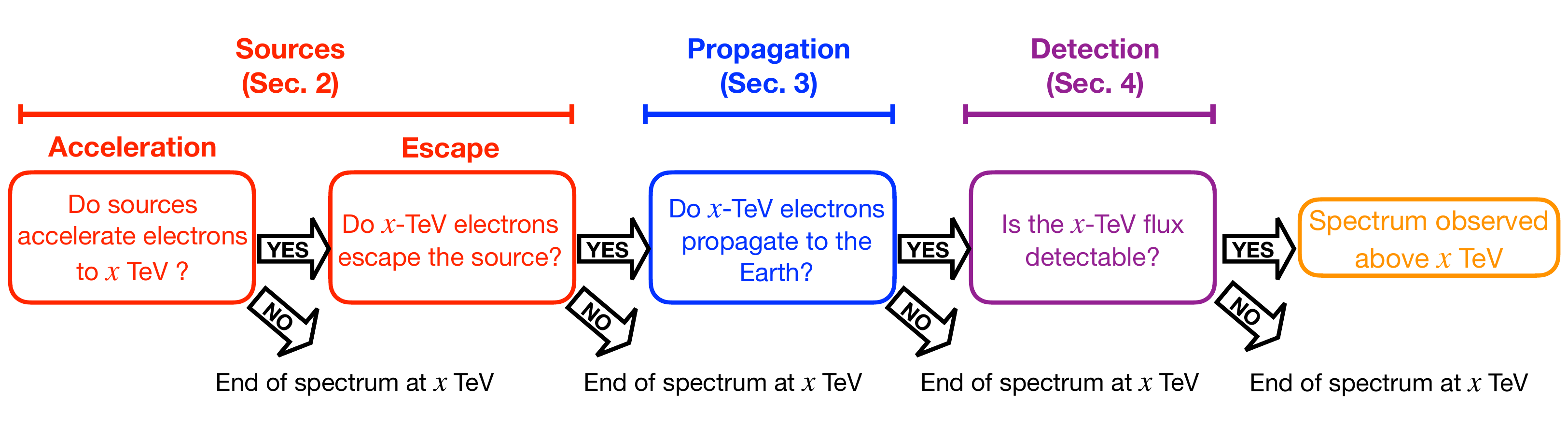}
    \caption{Flowchart to show the different steps (see the corresponding sections for details) that could each cause an end to the observed CR electron spectrum.  \textit{It is unknown where the spectrum ends and why.}}
    \label{fig:flow}
\end{figure*}

Figure~\ref{fig:flow} shows that the question of the end of the CR electron spectrum depends on four steps: (1) acceleration in the sources, (2) escape from the sources, (3) propagation from the sources to Earth, and (4) detection of the electrons.  We tackle these steps in an approximate but comprehensive way, focusing on establishing a framework to be explored in more detailed work.  For each of the four steps --- any of which could cause the end of the CR electron spectrum --- we show that existing data and theory allow a wide range of possibilities.  We outline strategies for resolving these sub-questions through multi-messenger observations and theory.

\begin{figure*}
    \centering    \includegraphics[width=2\columnwidth]{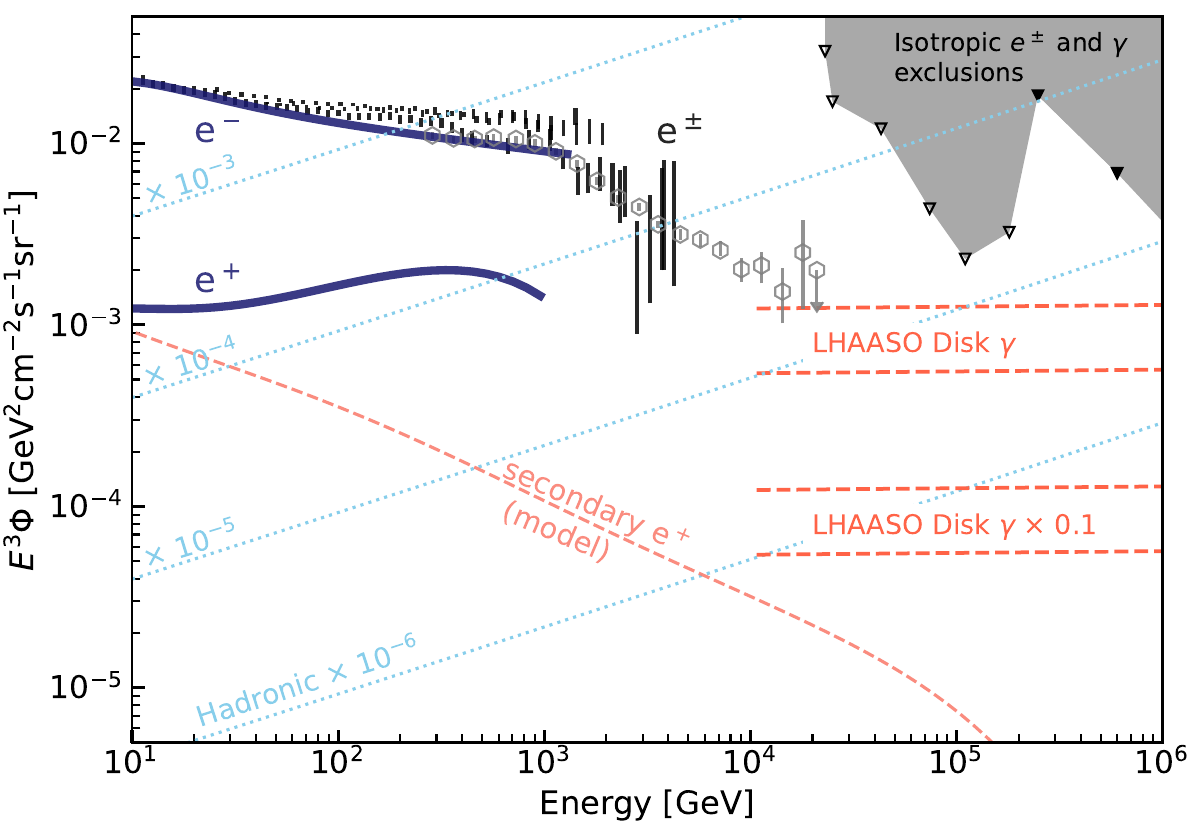}
    \caption{Overview of CR electron and positron observations.  For electrons ($e^-$ only; up to 1.4~TeV) and positrons ($e^+$ only; up to 1~TeV) we show with bold dark-blue curves the fits to AMS data \cite{2019PhRvL.122j1101A, 2019PhRvL.122d1102A}. For $e^- + e^+$ (denoted as $e^\pm$), we show with black bars the data from Fermi (up to 2~TeV~\cite{2017PhRvD..95h2007A}), CALET (up to 4~TeV~\cite{2018PhRvL.120z1102A}), and DAMPE (up to 4~TeV~\cite{2017Natur.552...63D}); the data from VERITAS~\cite{2018PhRvD..98f2004A} and MAGIC are consistent but are not shown. HESS results extend up to 20~TeV, but these data (grey open hexagons) are reported only in ICRC 2017 and still not yet in a publication (we take the data from a theory paper~\cite{2018PhRvD..98h3009H}).  At the highest energies, we show (gray shading) limits on the isotropic gamma-ray and electron flux from HAWC~\cite{2022arXiv220908106H} (open triangles) and KASCADE~\cite{2017ApJ...848....1A} (filled triangles; the limits are given relative to the cosmic-ray intensity, for which we use the fit in Ref.~\cite{2018PhRvD..98d3003L}. See also Ref.~\cite{2017PhRvD..96b3015Z}, which used other experiments' data, including preliminary data).  We also show relevant scales that may set a floor to the sensitivity of CR electron detection (details in Sec.~\ref{sec:detection}).  \textit{The CR electron spectrum is effectively unprobed at the highest energies.}
    }
    \label{fig:data}
\end{figure*}

Figure~\ref{fig:data} quantifies the observational status and prospects for the detection of CR electrons at the highest energies.  We show direct measurements of and limits on the CR electron spectrum, along with a rough sense of how the experimental sensitivity is limited by the background due to CR hadrons.  In addition, we show relevant scales that may determine how astrophysical foregrounds (e.g., due to gamma rays) limit the sensitivity.  The details of our calculations are given in Sec.~\ref{sec:detection}, where we also discuss the sensitivity of powerful new experiments, especially the Large High Altitude Air Shower Observatory (LHAASO)~\cite{2019ICRC...36..471W, LHAASO:2019qtb}.

In Sec.~\ref{sec:sources}, we discuss CR electron acceleration in and escape from sources. In Sec.~\ref{sec:propagation}, we discuss CR electron propagation. In Sec.~\ref{sec:detection}, we discuss their detection. Each of these three sections starts with general considerations, followed by estimates of the ranges of possibilities. In Sec.~\ref{sec:future}, we discuss multi-messenger strategies. In Sec.~\ref{sec:summary}, we conclude.

\clearpage
\section{Sources}
\label{sec:sources}

In this section, we discuss possible sources of TeV--PeV electrons, focusing on acceleration and escape.  While we show results for varying propagation models in Sec.~\ref{sec:propagation}, here we fix this to focus on source properties.  We use isotropic diffusion with $D_{\rm iso} = 3 \times 10^{29}$~cm$^2$~s$^{-1}$~$(E/\rm TeV)^{1/3}$, a magnetic field strength of 3~$\mu$G, and the radiation fields from the cosmic microwave background (2.7~K, 0.26~eV~cm$^{-3}$) and the local infrared background (20~K, 0.2~eV~cm$^{-3}$).  We take into account the Klein-Nishina effect following Ref.~\cite{2014ApJ...783..100K}.

Here and in subsequent sections, our calculations are approximate.  We seek to elucidate the range of possibilities, showing where more detailed calculations --- especially extending to the PeV range, which is not typical --- would be important to reducing uncertainties.


\begin{figure*}[t]
    \centering    \includegraphics[width=2\columnwidth]{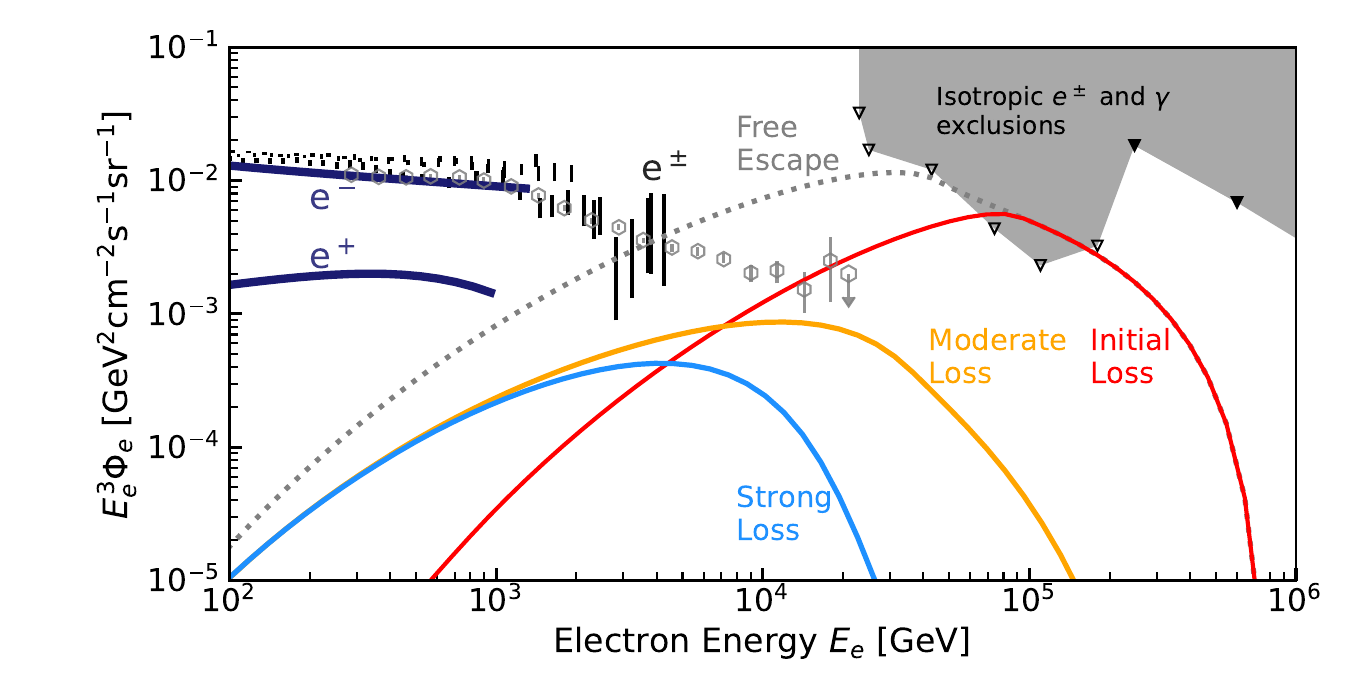}
    \caption{For a single pulsar, variations in the CR electron spectrum at Earth in various escape scenarios.  Where the predictions for a single source are above the data, such scenarios would be ruled out.  Where the predictions for a single source are below the data, additional studies would be needed to assess if the scenario can reproduce observations as the sum of multiple sources.  \textit{As the preliminary HESS results (grey open hexagons, same as in Fig.~\ref{fig:data}) could have a huge impact on constraining models, it is critical to test those results.}}
    \label{fig:escape-models}
\end{figure*}

\subsection{General Considerations}

In the GeV range, supernova remnants (SNRs) are believed to be the dominant sources of the CR electron flux. SNRs produce electron ($e^-$) primaries by directly accelerating them as well as positron and electron secondaries through the $pp$ interactions of accelerated CR hadrons.  The global fit to the GeV-range CR $e^-$ flux allows us to determine the total energy per source emitted in nonthermal electrons ($\mathcal{E}_{e,\rm SNR}\simeq 10^{48}$~erg) and the typical spectral index ($\gamma_{e,\rm SNR}\simeq 2.6$)~\cite{2021PhRvD.103h3010E, 2022ApJ...926....5A}. Present data suggest that the injected $e^-$ spectrum from SNRs may have a cutoff near 10~TeV, though this is uncertain~\cite{2022ApJ...926....5A}. 

In the TeV range, young and middle-aged pulsars, which produce both electron and positron primaries, may start to be important.  The total GeV flux from pulsars must be much smaller than that from SNRs because pulsars inject approximately equal numbers of $e^-$ and $e^+$ and the observed positron flux is about an order of magnitude smaller than the electron ($e^-$) flux. Nevertheless, at high energies, the pulsar contribution can be important, due the long lifetimes of pulsars and the hard spectra they produce.  It is well established that the maximum energies in such sources reach extreme energies.  An iconic young ($\simeq$1~kyr) pulsar, the Crab, produces PeV gamma rays~\cite{2021Sci...373..425L}, which indicates that electrons of even higher energies are present. Another well-studied pulsar, Geminga, produces electrons above about 100~TeV~\cite{2017Sci...358..911A}, despite its relatively old age  ($\simeq$300~kyr). Together with a variety of X-ray and gamma-ray observations, these observations indicate that pulsars can produce $>$100~TeV electrons over a time as long as 300~kyr (and possibly even longer). 

The total energy released by a pulsar over its lifetime is $\mathcal{E}_{\rm pul}\sim 10^{49}(P_0/\rm 50~ms)^{-2}$~erg ($P_0$ is the initial pulsar spin period), with electrons carrying $\sim$10\% of this~\cite{2019PhRvD.100d3016S}.  As the pulsar power decreases over time as $\propto(1+t/\tau_{\rm sd})^{-2}$, most of the power is released before the spindown timescale, which is $\tau_{\rm sd}\sim 4(P_0/\rm 50~ms)^{2}$~kyr. The distribution of $P_0$ is uncertain, with possible average values ranging between $\sim$50--300~ms~\cite{2006ApJ...643..332F, 2008ApJ...678L.113D, 2011ApJ...727..123W, 2012Ap&SS.341..457P, 2013MNRAS.432..967I, 2013MNRAS.430.2281N}. The spectral shape of electrons from pulsars is uncertain; above $\sim$1~TeV, the spectral index is usually constrained to be in the range $\gamma_{e,\rm pul}\simeq$2.0--2.5, though $\gamma_{e,\rm pul}\simeq1.5$ has also been commonly assumed for Geminga~\cite{2011ApJ...741...40T, 2014JHEAp...1...31T, 2017ApJ...835...66P, 2017PhRvD..96j3013H, 2018ApJ...863...30F, 2018PhRvD..97l3008P, 2019MNRAS.484.3491T}.

What is more uncertain and more important is the spectrum of {\it escaped} electrons. This has gained renewed attention since the discovery of ``TeV halos" around pulsars~\cite{2019PhRvD.100d3016S, 2007ApJ...664L..91A, 2017Sci...358..911A, 2017PhRvD..96j3016L, 2020A&A...636A.113G, 2022NatAs...6..199L, 2022FrASS...922100F, 2022IJMPA..3730011L}, which indicate that electrons do escape from the shocked regions in the pulsar wind nebulae but remain efficiently confined in the source vicinity. The degree of confinement by TeV halos is under debate. Geminga observations are commonly interpreted as indicating that the diffusion coefficient in the source vicinity is suppressed from its typical galactic value by a factor of $\sim$100--1000, in which case electrons above $\sim$10~TeV would lose their energy in the halo (for a size of 50~pc). However, alternative models with no suppression have been proposed~\cite{2021PhRvD.104l3017R} (though these models are disputed~\cite{Bao:2021hey, Mukhopadhyay:2021dyh, Bi:2023nxr}), in which case particles of $>$1~PeV could escape. Even if the strong suppression is true for Geminga, it is a separate and unresolved question if this is ubiquitous~\cite{2022A&A...665A.132M}.  If so, as seems to be favored, this could have interesting aggregate effects on CR propagation~\cite{2022ApJS..262...30P, Zhao:2021yzf, 2023MNRAS.526..160J}; below, we comment on the possible impacts for CR electrons.

Due to CR electron cooling during propagation (discussed in Sec.~\ref{sec:propagation}), only sources within approximately 700~pc can be relevant for CR electron observations above 10~TeV. Based on Refs.~\cite{2004ApJ...601..340K, 2014JCAP...04..006D}, the following sources are often considered as promising: \textit{G114.3+00.3} (700~pc, 7.7~kyr), \textit{Vela Jr.} (750~pc, 1.7--4.3~kyr; we use 3~kyr), \textit{Cygnus Loop} (440~pc, 20~kyr), \textit{Monogem} (300~pc, 86~kyr, associated with  PSR B0656+14), \textit{Vela~YZ} (300~pc, 11~kyr, associated with PSR B0833-45), \textit{Loop~I} (170~pc, 200~kyr), and \textit{Geminga} (250~pc, 330~kyr, associated with PSR J0633+1746). (See also Ref.~\cite{2019ApJ...884..124F}, which discusses the possible importance of PSR B1055-52.) We quote the distances and ages from Refs.~\cite{2004ApJ...601..340K, 2014JCAP...04..006D}, except for Geminga, for which we use Ref.~\cite{2012ApJ...755...39V}.

Sources besides those listed above can also be important. Based on the ATNF pulsar catalog~\cite{2005AJ....129.1993M}, there are 17 pulsars within 700~pc that are younger than 1~Myr. These are only a minor fraction of the total population; due to the the pulsar emission being beamed, only $\sim$25\% of pulsars are visible~\cite{1998MNRAS.298..625T}, so there could be $\sim$50 more that remain undiscovered. Their parent SNRs may be difficult to detect if their age significantly exceeds 100~kyr, which is when SNRs start to blend into the interstellar medium (for example, the SNR for Geminga is not observed). In most previous studies, under the assumption of isotropic diffusion, such ``unnamed" sources are thought to be unimportant compared to the above-listed, bright, and well-studied objects. (For an exception, see, e.g., Ref.~\cite{2018PhRvL.121y1106L}.) However, diffusion may be strongly anisotropic, with particles preferentially propagating along magnetic field lines (see Sec.~\ref{sec:propagation}). In such scenarios, a random source that happens to be close to a local magnetic field line can be significantly more important than sources that are more luminous.

Other class sources may also contribute to the CR electron and positron fluxes. In particular, millisecond pulsars could be important~\cite{2012MNRAS.421.3543K}, because various observations point to the production of nonthermal particles by millisecond pulsars~\cite{2021MNRAS.507.5161S, 2022NatAs...6.1317C, 2022MNRAS.516.4469Z, 2022PhRvD.105j3013H}. They are individually dim compared to the young pulsar sources discussed above, but are more numerous, and hence possibly make a sizable contribution to the observed flux. Sources like Galactic black holes~\cite{2017MNRAS.470.3332I, 2019MNRAS.488.2099T, 2022ApJ...935..163S}, white-dwarf pulsars~\cite{2011PhRvD..83b3002K}, or exotic sources like dark matter~\cite{2005PhR...405..279B, 2012JCAP...10..043M, 2017PhRvL.119b1102C, 2019PhRvD.100d3029S, 2021JCAP...12..007J, 2022PhRvL.129z1103C, 2023PhRvD.107l3027Z} could also be relevant.


\subsection{Range of Specific Possibilities}

Here, we illustrate that models for particle escape from pulsars are crucial to where and how the electron spectrum ends. To do so, we solve the one-dimensional diffusion-loss equation with the method of Ref.~\cite{2022ApJ...926....5A} (which builds on Ref.~\cite{1995PhRvD..52.3265A}) to obtain the density of electrons near Earth, $n_e$. The observed electron intensity (flux per solid angle) is then $\Phi_e = cn_e/(4\pi)$.

While this method is standard, two points need attention. First, we should propagate particles up to the {\it actual} age, $T_{\rm age}$, rather than the {\it observed} age, $t_{\rm age, obs}$, because what appears in the diffusion-loss equation as $t$ is the physical time. As the observed age is estimated based on electromagnetic observations, we expect $T_{\rm age} = t_{\rm age, obs} + R/c$, where $R$ is the distance to the source. Second, the ordinary diffusion equation has a ``superluminal propagation" problem. This can be an issue in particular for the case of continuous injection: no particles that are injected between $t_{\rm age, obs}$ and $T_{\rm age}$ should arrive at the Earth, while some do in the naive diffusion approximation. While phenomenological solutions exist in the literature~\cite{2009ApJ...693.1275A}, this is not yet extended to include the case of anisotropic diffusion, which we discuss below.  Here, we avoid the superluminal propagation problem by simply setting the injection term to zero between $t=t_{\rm age, obs}$ and $t=T_{\rm age}$.  Though this treatment does not entirely remove the problem, it does remove particles that are a priori too young and should not arrive at Earth, while keeping other particles unaffected.

As an example electron source, we consider a pulsar at 440~pc and 20~kyr (similar to the Cygnus Loop), and fix its total electron energy (for particles above 1 GeV) to $10^{48}$~erg and assume a spindown timescale of 10~kyr. For the spectrum of electrons that escape into the interstellar medium, we estimate results for the following scenarios:
\begin{itemize}

    \item
    {\it Free Escape}: Particles easily escape from the source, and the escaped spectrum of electrons is similar to the accelerated spectrum.  We assume $\gamma_{e, \rm pul}=2.3$ and set a cutoff energy $E_{e,\rm cut}=1$~PeV. Note that the spectral index might be even harder.

    \item
    {\it Initial Loss}: Particles can escape the source only after the initial evolution of the pulsar wind nebula. This may occur if the magnetic field is strong and ordered when the source is young (like the Crab), effectively confining electrons and leading to strong synchrotron losses. To estimate this effect, we assume that no particles produced in the initial 10~kyr of the pulsar's lifetime can escape the source region. After 10~kyr, the particle spectrum is set to be the same as for the Free Escape case.

    \item
    {\it Moderate Loss}: Particles lose some energy before escaping the source region. The impact of energy loss at the source has been extensively studied using multi-zone models (e.g., Ref.~\cite{Schroer:2023aoh}); generally, the escaping spectrum is softer than the accelerated spectrum.  We assume a slightly softer index ($\gamma_{e, \rm pul}=2.5$) than above and a smaller cutoff ($E_{e,\rm cut}=100$~TeV). Note that the spectral index might be even softer.

    \item
    {\it Strong Loss}: Particles are efficiently confined by the TeV halo, losing much of the energy. We assume $\gamma_{e, \rm pul}=2.5$ and even smaller cutoff $E_{e,\rm cut}=10$~TeV.

\end{itemize}

Figure~\ref{fig:escape-models} shows that the end of the electron spectrum is very sensitive to the details of the particle-escape model. Even for a single pulsar, the ``Free Escape" scenario could be strongly in tension with the preliminary HESS results above 5~TeV and mildly in tension with the HAWC limits near 100 TeV.  For the ``Initial Loss" scenario, the tension with HESS is removed but that with HAWC remains, also even for a single pulsar.  For both the ``Free Escape" and the ``Initial Loss" scenarios, future LHAASO measurements at very high energies will be of crucial importance. Note that the spectral index for these two cases might be even harder, in which case the tension would be more severe. For the ``Moderate Loss" scenario, the contribution from this single pulsar would fall below all measurements, and the CR electron spectrum would end smoothly. However, it needs to be asked if the contributions from multiple sources would make make the overall spectrum too large; in the next section we show that above $\sim$10~TeV, typically only one or at most a few pulsar contributes, suggesting that the ``Moderate Loss" scenario is likely allowed even if multiple pulsars are considered. Finally, for the ``Strong Loss" scenario, the contribution from this pulsar would fade below $\sim$10~TeV, making pulsars of little importance near the end of the electron spectrum. In this case, SNRs would likely be more important sources than pulsars all the way from GeV energies to the end of the electron spectrum.

As a next step, more detailed theoretical efforts will be important to investigate the impact of particle escape. Such studies have been carried out in, e.g., Refs.~\cite{2011ApJ...729...93K, 2012MNRAS.427...91O}, though their attention is typically at lower energies than we discuss here. Focusing on the end of electron spectrum would be fruitful because it is very sensitive to the source model, as we show above. 


\section{Propagation}
\label{sec:propagation}

In this section, we discuss the propagation of TeV--PeV electrons for fixed source models.  For SNRs, we assume that electrons are injected into the interstellar medium with a spectrum $dN_e/dE_e\propto E_e^{-2.6}\exp(-E_e/E_{e,\rm cut})$, normalizing it to $10^{48}$~erg above 1~GeV and using $E_{e,\rm cut}=30$~TeV.  For pulsars, we use ``Moderate Loss" model above, again fixing the integrated electron energy to $10^{48}$~erg. Again, our focus is on approximate calculations that show the range of possibilities.


\subsection{General Considerations}

Once accelerated by and escaped from sources, charged particles propagate in the galaxy's $\sim$$\mu$G interstellar magnetic field, where they scatter with turbulence, leading to diffusive CR propagation.  The nominal Larmor radius is $r_L \simeq 1$~pc~$(E/{\rm PeV})(B/\rm \mu G)^{-1}$ and the coherence length of the turbulence is $l_{\rm coh}\sim 1$--30~pc~\cite{2008ApJ...680..362H, 2010ApJ...710..853C, 2013A&A...558A..72I, 2018MNRAS.479.4526L}, depending on the location in the Milky Way.  The physical scales are obtained with phenomenological transport models fit to various primary and secondary CR data; the obtained diffusion coefficient is $D_{\rm iso} \simeq 3 \times 10^{29}$~cm$^2$~s$^{-1}$~$(E/\rm TeV)^{1/3}$~\cite{2010ApJ...722L..58S, 2017JCAP...02..015E, 2022ApJS..262...30P, 2019PhRvD..99j3023E, 2023ApJ...949...16S}, which corresponds to a mean free path of $l_{\rm mfp} \sim 10$~pc~$(E/\rm TeV)^{1/3}$.  Below, we consider sources much more distant than $l_{\rm coh}$ and $l_{\rm mfp}$, so that diffusive approximation is valid.

\begin{figure}[b]
    \centering
    \includegraphics[width=\columnwidth]{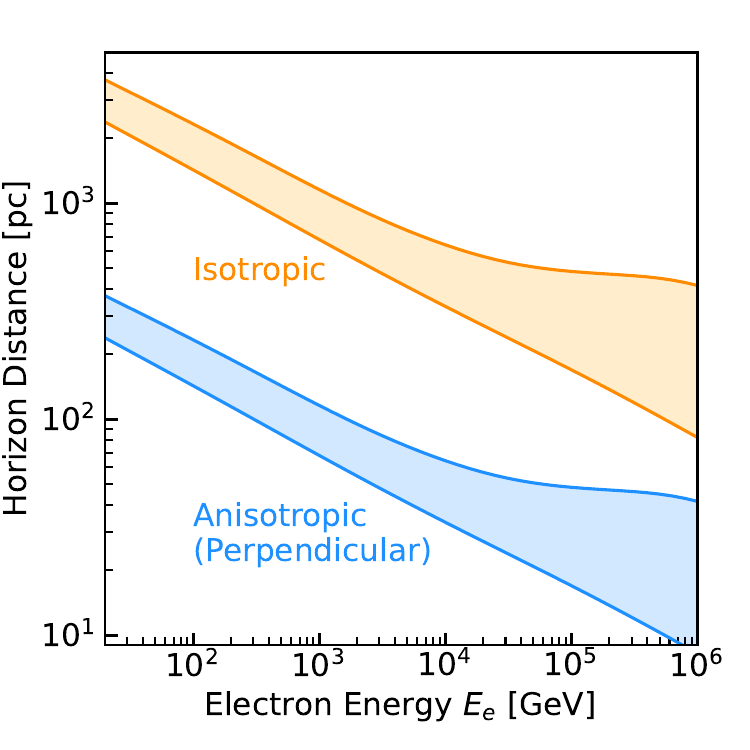}
    \caption{Horizon distances for CR electrons in the isotropic and anisotropic diffusion cases, as labeled, with the bands showing the effect of varying the magnetic field strength from 1~$\mu$G (upper) to 6~$\mu$G (lower).  For the anisotropic case, we take $D_\parallel = D_{\rm iso}$ and $D_\perp = D_\parallel/100$.  \textit{The horizon distance for observable CR electron sources is small, especially at very high energies.}}
    \label{fig:horizon-length}
\end{figure}

The horizon distance an electron can travel strongly depends on the propagation model.  The underlying reason is that the diffusion and cooling of electrons limit the distance they can travel (the ``horizon distance") to $l_{\rm cool} \sim \sqrt{2Dt_{\rm cool}}$, which is only less than 700~pc above 10~TeV (see below) and decreases with energy due to $t_{\rm cool} \sim 60(E/\rm 10~TeV)^{-1}$~kyr. To become relevant, a source needs to be located within a volume of $V_{\rm cool}^{\rm (iso.)}\sim \pi l_{\rm cool}^3$ (for isotropic diffusion). As the particle energy increases, the volume shrinks, decreasing the number of sources inside it.  Though we use the global diffusion coefficient from CR hadron data, the local diffusion coefficient near Earth --- needed for CR electrons --- could be different.

While isotropic diffusion is often assumed, anisotropic diffusion may be more appropriate.  Isotropic diffusion is appropriate if particles experience a number of random field configurations, averaging out the CR directions. However, the Milky Way's large-scale, ordered magnetic field may have a significant impact on CR propagation. The role of anisotropic diffusion has been intensively studied for CR hadrons~\cite{2012A&A...547A.120E, 2012PhRvL.108u1102E, 2014ApJ...785..129K, 2015PhRvL.114b1101M, 2016PhRvL.117o1103A, 2017PrPNP..94..184A, 2019JPhCS1181a2039K}. It should be even more critical for CR electrons because their propagation distances are smaller and the effects can be large~\cite{2012PhRvL.108z1101G, 2012arXiv1210.8180K, 2021JCAP...02..030L}.

Anisotropic diffusion occurs when particle transport is more efficient parallel to the magnetic field lines than perpendicular~\cite{2008ApJ...673..942Y, 2020SSRv..216...23S, 2021ApJ...923..209S}, as CRs will travel in helical trajectories along the field lines as opposed to isotropic random walks.  The ratio $D_\perp/D_\parallel$ could be as small as $\sim$1/1000--1/100 and might be energy-dependent~\cite{2001PhRvD..65b3002C, 2007JCAP...06..027D, 2010ApJ...725.2117S,  2018JCAP...07..051G, 2022SNAS....4...15R, 2022arXiv221105881K, 2022arXiv221105882K}. Correspondingly, the horizon distance perpendicular to the field, $l_{\perp}$, is smaller than the parallel one, $l_{\parallel}$, by a factor of $\sqrt{D_\parallel/D_\perp}$.  Anisotropic diffusion thus reduces the CR electron horizon volume to $V_{\rm cool}^{\rm (aniso.)}\sim \pi l_{ \parallel}l_{\perp}^2$, where these separate cooling lengths are defined analogously to that in the isotropic case.  To obtain the same CR flux at Earth, a smaller number of sources must then contribute brighter fluxes.

\begin{figure*}[t]
    \centering
    \includegraphics[width=2\columnwidth]{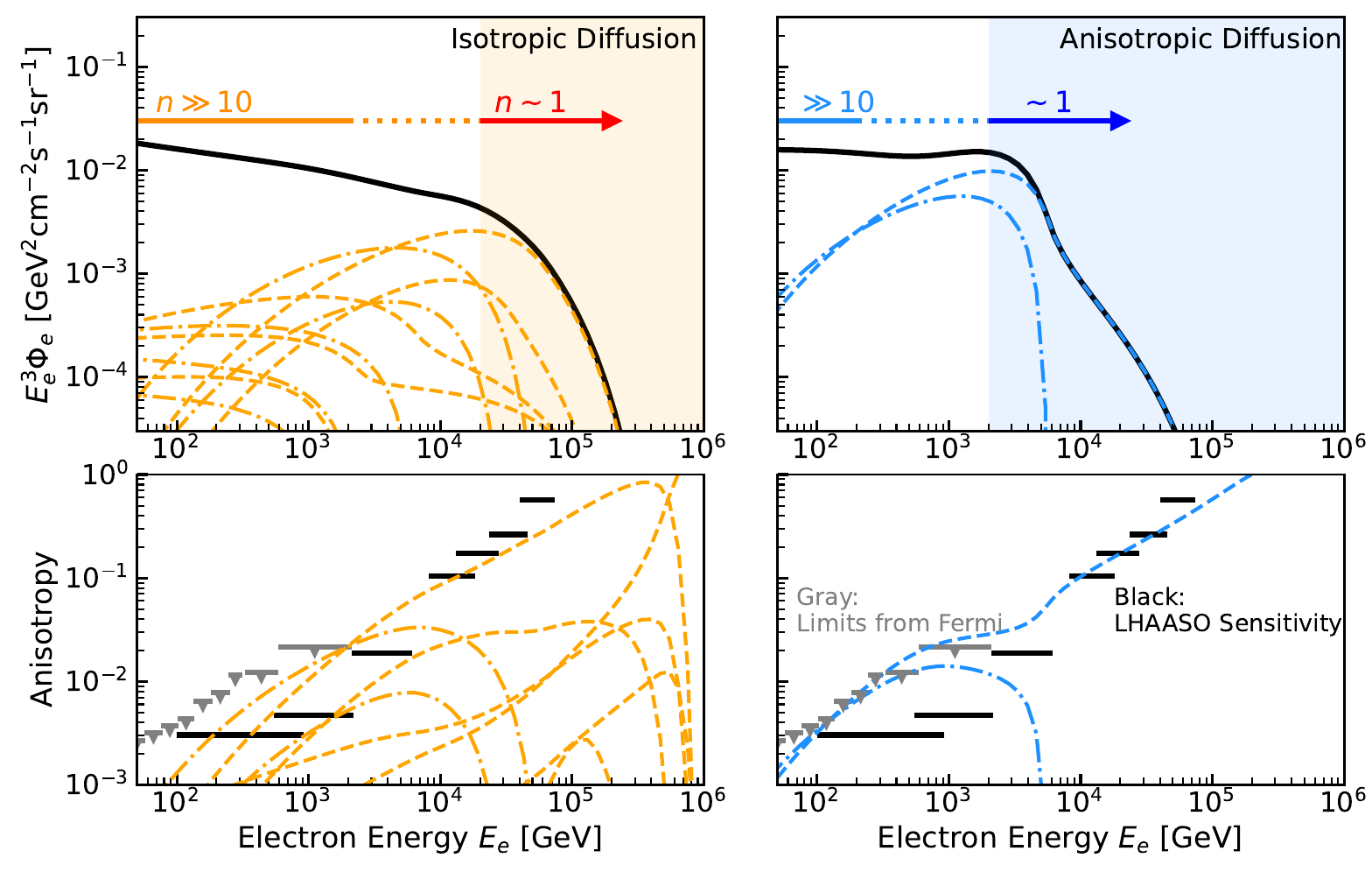}
    \caption{How the composite CR electron spectrum (upper panels) and source dipole anisotropies (lower panels) depend on the propagation model (isotropic diffusion in the left panels, anisotropic diffusion in the right panels).  The dash-dotted lines show contributions from individual SNRs, while dashed lines show those from individual pulsars. Shaded areas show the approximate energy ranges where a single source could dominate. \textit{A key test for the diffusion models is the energy range where only a single source contributes and then the spectrum plummets.}}
    \label{fig:horizon-prop}
\end{figure*}

Figure~\ref{fig:horizon-length} shows how the horizon distance can depend on the propagation model (isotropic or anisotropic) as well as the uncertain magnetic field strength. In this figure and below, we assume $D_\parallel = D_{\rm iso}$ (required to be consistent with hadronic CR data) and $D_\perp = D_\parallel/100$. For the isotropic case, we can see that the sources must be nearby; above 10~TeV, the horizon distance is below 700~pc. For the anisotropic case, the source locations are even more strongly constrained; above 10~TeV, the horizon distance perpendicular to the magnetic field is less than 100~pc, meaning that the source has to be very close to a magnetic field line that passes near Earth.

The number of sources that can contribute to the observed CR electron flux is determined by the horizon volume.   We estatime the expected source count, $n$, as 
\begin{equation}
    n \sim V_{\rm cool} / V_{\rm src},
\end{equation}
where $V_{\rm src}$ is the typical volume in which one source is contained.  To contribute, a source need not only be within $V_{\rm cool}$; it must also inject CR electrons within a time $t_{\rm cool}$ before the present.  Therefore, the source age and electron cooling time must be accounted for in estimating $V_{\rm src}$. For SNRs, this means that the ages need to be younger than $t_{\rm cool}$, so then the source volume can be estimated as $V_{\rm src} \sim V_{\rm MW}/\Gamma_{\rm snr}t_{\rm cool}$, where $V_{\rm MW}$ is the volume of the Milky Way, for which we assume $2\pi$(15~kpc)$^2$(500~pc), and $\Gamma_{\rm snr}$, for which we assume 0.03~yr$^{-1}$, the Galactic core-collapse supernova rate~\cite{2013ApJ...778..164A}. For pulsars, the volume is estimated using the pulsar birth rate $\Gamma_{\rm pul}$, which we assume to be the same as $\Gamma_{\rm snr}$, and the lifetime of pulsars as electron sources (i.e., how long a pulsar can inject electrons after their birth), $t_{\rm inj}$.  We thus obtain $V_{\rm src} \sim V_{\rm MW}/\Gamma_{\rm pul}(t_{\rm cool}+t_{\rm inj})$. Note that $t_{\rm inj}$ could be energy dependent (i.e., pulsars might produce PeV electrons only when they are young, while producing lower-energy electrons over a much longer time.) 

For the example of SNR sources, we find that the numbers of contributing sources are quite different for the cases of isotropic and anisotropic diffusion:
\begin{equation}
    n^{\rm (iso.)} \sim 100\left(\frac{\rm TeV}{E}\right)^2,
\end{equation}
and 
\begin{equation}
    n^{\rm (aniso.)} \sim 1\left(\frac{\rm TeV}{E}\right)^2.
\end{equation}
The energy dependence is due to $V_{\rm cool} \propto E^{-1}$ and $V_{\rm src} \propto t_{\rm cool}^{-1} \propto E$.  For pulsars, we expect that the energy dependence might be weaker because $t_{\rm cool}\ll t_{\rm inj}$, so that $V_{\rm src}$ may not increase as rapidly as $\propto E$. For more careful estimates, we would need to take the geometry of the Milky Way disk into account, but doing so would not change our general points. 

The above calculation reveals three energy ranges for how many sources contribute to the observed CR electron flux:
\begin{itemize}

    \item
    Continuum ($n\gg10$): A large number of sources contribute, leading to a smooth spectral shape. The relevant energy range is $E\ll$~3~TeV for isotropic diffusion and $E\ll$~0.3~TeV for anisotropic diffusion.
    
    \item
    Transition ($n\sim$1--10): At higher energies, the spectrum may show fluctuations at the tens-of-percent level due to individual source contributions not averaging out.
    
    \item
    Individual ($n\sim$1): Statistically, we expect only a single source. The relevant energy range is $E\sim$~10~TeV for isotropic diffusion and $E\sim$~1~TeV for anisotropic diffusion. In this regime, statistical fluctuations are dominant, and the expected source may be present or not.
    
\end{itemize}
These estimates take into account only propagation effects.  If only a fraction of sources in a given class can accelerate electrons that then escape, then the numbers would be lower.


\subsection{Range of Specific Possibilities}

Here, we show that models for particle propagation are crucial to where and how the CR electron spectrum ends.  We calculate results for the following two scenarios:
\begin{itemize}

    \item
    {\it Standard Isotropic:} We apply the same methods and parameters as in Sec.~\ref{sec:sources} to calculate the fluxes from individual sources, focusing on the seven specific sources listed there.  We assume that each SNR has an associated pulsar, even if there is none reported. On top of the individual fluxes, we add a continuum of various sources with a cutoff at 3~TeV, normalizing the intensity and spectrum from the observed GeV CR $e^-$ data.
  
    \item
    {\it Strong Anisotropic:} We estimate the flux as in the ``Standard Isotropic" case, but increases it by hand by a factor of $D_\parallel/D_\perp$, which we assume to be 100; this appropriately takes the impact of anisotropic diffusion as long as $D_\parallel/D_\perp$ is energy independent and $r_\perp/l_\perp \ll 1$, where $r_\perp$ is the distance perpendicular to the magnetic field. In the presence of strong anisotropic diffusion, sources that appear promising may make little contribution, while sources that are otherwise unremarkable could be important. Here, for illustrative purposes, we assume a single system of a SNR + pulsar of age 100~kyr and distance 800~pc. We add a continuum component, as above, but with a lower cutoff of 0.3~TeV. 
    
\end{itemize}

Figure~\ref{fig:horizon-prop} (top panels) shows our calculations for the CR electron spectra at Earth for the cases of isotropic and anisotropic diffusion. For the isotropic case, the spectrum is smooth up to energies $>$1~TeV. In the regime where the source count is $n\sim$1--10, fluctuations are present, though they are small because each source has a broad spectrum. Note that while our results broadly agree with the existing data, they do not reproduce some spectral features, including the possible drop near 1~TeV; physical mechanisms not included in this calculation (e.g., electrons from SNRs may have a small cutoff energy) would be needed to explain such features. At the highest energies, Vela dominates the end of the spectrum.  For the anisotropic case, the dominance of a single source starts at $\sim$1~TeV. This means that the observed electron flux is already reflecting the end of the electron spectrum and that the spectral shape is determined by this single source.

Figure~\ref{fig:horizon-prop} (bottom panels) shows the expected source dipole anisotropies.  To obtain the total anisotropy, we would have to add these with vectors, which would require the locations of each source. At very high energies, where $n\sim$1, the expected anisotropy can be large.  When the anisotropy approaches the maximum value of unity, its calculated value may not be completely accurate due to the superluminal diffusion problem noted above; a more careful treatment is needed.

As a next step, more detailed theoretical efforts will be important to deepen the knowledge on the role of anisotropic propagation. This would not only be crucial to where the CR electron spectrum ends, but also be important to understand the CR of all particles, including hadrons, as well as diffuse emission of gamma rays and neutrinos~(see, e.g., Refs.~\cite{2017JCAP...10..019C, 2023arXiv230510251G}).

Last, we remark on the possible effects of TeV halos on the propagation of CR electrons.  As a general point, we expect that two-zone diffusion models~\cite{2018ApJ...863...30F, 2018PhRvD..97l3008P, 2019MNRAS.484.3491T} --- with slow diffusion around sources and normal diffusion in between --- will be needed, especially at the highest CR electron energies.  Here, where our focus is on the big picture, we simply remark on some possible scenarios, deferring the many complexities to future work.  In one scenario, Earth is inside a TeV halo; this case is like isotropic (or possibly anisotropic) diffusion with a much lower diffusion coefficient, so that the source count should be small, and the cutoff energy should be low.  In another possible scenario, Earth is in a normal-diffusion region, relatively far from any TeV halo (or they are more compact); for this case, we expect something like the ``Strong Loss" model, where electrons lose most of the energy before leaving the TeV halo.  To consider these and other scenarios in detail, better observations of TeV halos with LHAASO~\cite{LHAASO:2019qtb}, the Cherenkov Telescope Array (CTA)~\cite{2019scta.book.....C}, and other detectors will be critical.


\section{Detection}
\label{sec:detection}

In this section, we assess some key factors for the detectability of the CR electron spectrum at the highest energies, focusing on detector backgrounds and astrophysical foregrounds.  While detailed assessments of sensitivity must be done by the experimental collaborations, our results show that significant discovery space seems to be within reach.


\begin{figure*}[t]
    \centering    
    \includegraphics[width=2\columnwidth]{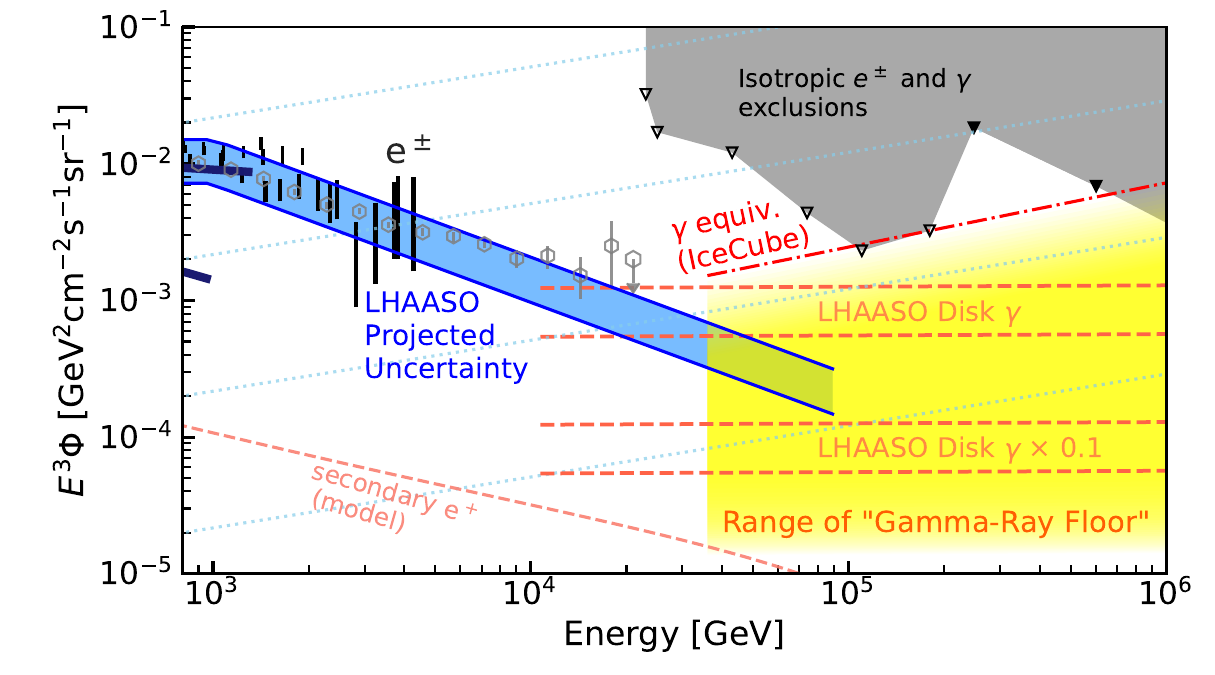}
    \caption{Detection prospects for CR electrons at the highest energies.  Starting with a zoomed-in version of Fig.~\ref{fig:data} with some labels omitted, we add the ``Projected Uncertainty" band from LHAASO (which assumed that the spectrum follows the trend of the HESS preliminary results)~\cite{2019ICRC...36..471W}.  The yellow region shows our estimate of the range for the CR-electron detection ``floor" due to the quasi-isotropic component of the Milky Way's gamma-ray emission. Note that the x-axis refers to either electron or gamma-ray energy. \textit{At the highest energies, probing CR electrons requires doing the same for Milky Way gamma rays.}}
    \label{fig:floor-models}
\end{figure*}

\subsection{General Considerations}

Towards the PeV range, CR electrons must, due to their small fluxes, be detected through the electromagnetic showers they induce in Earth's atmosphere.  In the lower TeV range, the best sensitivity has been through ground-based air-Cherenkov detectors that register only the light from the showers.  At higher energies, the best sensitivity has been through ground-based arrays that register the shower particles themselves.

As we focus here on the highest energies, the most relevant experiment is LHAASO~\cite{2019ICRC...36..471W, LHAASO:2019qtb}.  However, we note that significant improvements in sensitivity are expected at intermediate energies and that such new measurements will be important for better understanding the origins of the electron spectrum at all energies.  Direct (space-based) measurements of the electron spectrum will be extended to beyond tens of TeV by the High Energy cosmic-Radiation Detection (HERD) experiment~\cite{2022PhyS...97e4010K, Berti:2023lah}.  And it will be possible to go well beyond the indirect (ground-based) measurements of HESS~\cite{2018PhRvD..98h3009H} in both energy range and precision with CTA~\cite{2019scta.book.....C}.

Figure~\ref{fig:floor-models} shows the uncertainties (dominantly systematics) expected for a LHAASO measurement of the CR electron spectrum~\cite{2019ICRC...36..471W}.  In this estimate, they have assumed that the CR electron spectrum continues as a power law that largely follows the preliminary HESS results.  To obtain this uncertainty band, LHAASO had to take into account the projected exposure of the detector as well as the expected detector backgrounds.  The results are quite encouraging for the detection of the CR electron spectrum at very high energies, even if the data does not follow the trend they assumed.

The most important background for CR electron detection is caused by CR hadrons, which also produce showers in Earth's atmosphere, though these hadronic showers have different morphological features and have accompanying muons. LHAASO can powerfully reject hadrons; they foresee hadron survival fractions of $\sim$$10^{-3}$ and $\sim$$10^{-5}$ at 10~TeV and above 100~TeV, respectively, while retaining more than half of CR electrons~\cite{2019ICRC...36..471W}. Despite such strong cuts, the level of background is still high, because the flux of CR hadrons is overwhelmingly larger the flux of CR electrons.  To illustrate this, in Fig.~\ref{fig:data} and Fig.~\ref{fig:floor-models}, we show the CR hadron flux, multiplied by the survival factors as marked.  To achieve even better hadron rejection power, LHAASO will also employ new techniques, based on boosted decision trees, to statistically measure the CR electron fraction, even if individual events cannot be classified with certainty.  

We point out that it is also important to consider astrophysical foregrounds, by which we mean other fluxes of CR electrons or effectively identical particles.  The most obvious example is secondary electrons, which are produced by CR hadrons interacting with gas and producing pions.  In Figs.~\ref{fig:data} and \ref{fig:floor-models}, we show the expected flux that we estimated with a standard leaky-box model assuming that energy loss dominates over escape for electrons~\cite{2019PhRvD..99d3005L}.  Our estimates are similar to those in Refs.~\cite{2018PhRvD..98h3009H, 2021PhRvD.103h3010E, 2023arXiv230401261D}, for which some results are larger (up to about a factor of 3 at 1~TeV) or smaller than ours.  In any case, secondary electrons produced in the interstellar medium are unimportant as a foreground.  Secondary electrons produced in dense sources could have a larger flux, but likely still well below the primary CR electron spectrum. 

Another foreground --- one that we have not seen considered in this context --- is astrophysical gamma rays, which induce electromagnetic showers in Earth's atmosphere that are virtually indistinguishable from those created by CR electrons.  Any quasi-isotropic gamma-ray flux would define a ``floor" below which it would be hard to improve sensitivity to the CR electron flux.  At the highest energies shown in Fig.~\ref{fig:floor-models}, extragalactic gamma rays are heavily attenuated but Milky Way gamma rays are only partially attenuated~\cite{2006ApJ...640L.155M, 2016PhRvD..94f3009V, 2017MNRAS.470.2539P}.  We detail the possibilities in the next subsection.

If CR electrons are detected at very high energies, it will be important to test for the presence of anisotropies, as mentioned in Sec.~\ref{sec:propagation}.  With many sources, the dipole anisotropies from each may average out but, as the energy increases, it is more likely that there is only a single source.  As shown in Fig.~\ref{fig:horizon-prop}, LHAASO's estimates~\cite{2019ICRC...36..471W} of their sensitivity to a dipole anisotropy are quite encouraging.


\subsection{Range of Specific Possibilities}

At TeV--PeV energies, the Milky Way's diffuse gamma-ray emission should be morphologically similar to that observed in the GeV range by Fermi, which found that while the brightest emission is from the disk, there is some emission from all directions.  The primary cause is pion-producing hadronic CR collisions with gas, where the intensity in a given direction is proportional to the product of the CR density, the gas density, and the length of the line of sight.  In making measurements of the isotropic CR electron signal, LHAASO will avoid the Milky Way plane and thus the brightest gamma-ray emission, but a quasi-isotropic component will remain at high latitudes.  The spectrum of this emission follows that of the CR hadrons, i.e., $\sim$$E^{-2.7}$.  It is possible that emission from sources could be increasingly important at very high energies, due to sources having harder spectra, e.g., $\sim$$E^{-2.2}$; some sources could be at moderately high latitudes.

\begin{figure*}[t]
    \centering
    \includegraphics[width=2\columnwidth]{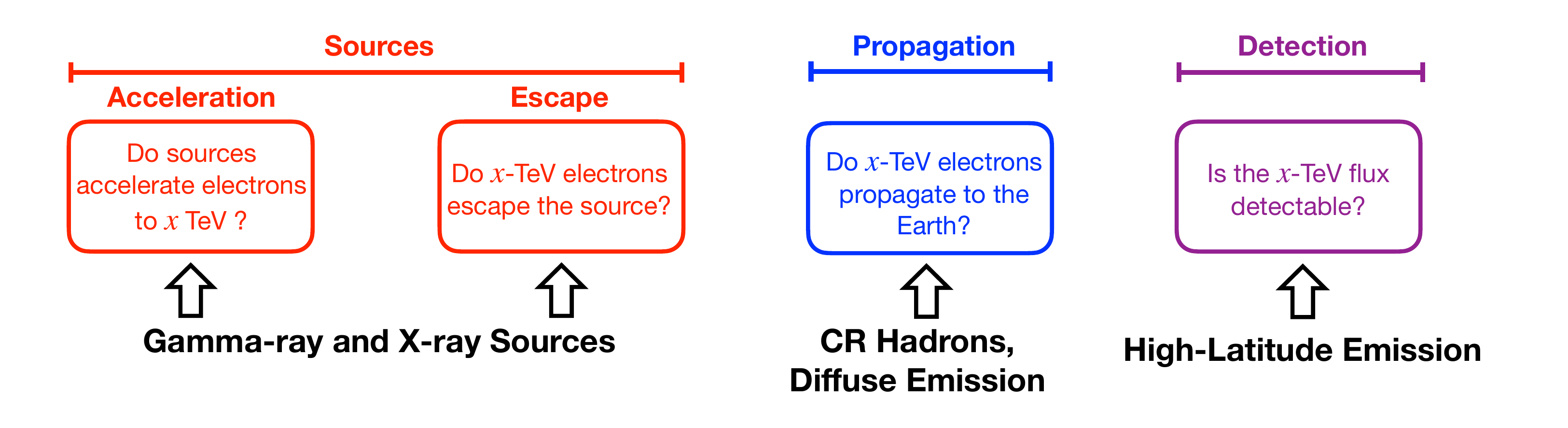}
    \caption{Multi-messenger approach towards understanding the end of the CR electron spectrum. \textit{While direct detection of CR electrons is critical, it will not be enough to separate the physics of these four steps.}}   
    \label{fig:future-MM}
\end{figure*}

Figure~\ref{fig:floor-models} shows, via the yellow shaded region, a broad range of possible intensity values for the gamma-ray floor.  As part of that, we show scales for guaranteed but uncertain diffuse gamma-ray emission.  The higher dashed band is set by LHAASO observations of the Milky Way plane, and the lower dashed band is set by a rough estimate of how much dimmer the high-latitude emission could be.  We estimate this component based on the Fermi data, for which the averaged high-latitude intensity ($|b|>20^\circ$) is smaller than the lower-latitude intensity by approximately a factor of 10~\cite{2012ApJ...750....3A, 2015ApJ...799...86A, 2023arXiv230506948Z}. As Fermi high-latitude emission has extragalactic component, which is absent in the LHAASO band, this treatment might slightly overestimate the floor.

In addition to these expected components of the gamma-ray emission, there is some possibility of surprises.  While IceCube's quasi-isotropic diffuse neutrino flux is likely extragalactic, some have argued that it could instead arise, at least in part, from emission in the Milky Way halo~\cite{2014PhRvD..89j3003T, 2018PhRvD..98b3004N, 2019PhRvL.122e1101B, 2021ApJ...914..135R} (See also Ref.~\cite{2020PhRvD.101l3023B} for a relevant model).  This could produce a quasi-isotropic diffuse gamma-ray flux as well that can be estimated by assuming that both the neutrinos and the gamma rays ultimately arise from pion decays.  This flux level also sets a scale, we show by a dot-dashed line in Fig.~\ref{fig:floor-models}.  Interestingly, the direct upper limits on the isotropic gamma-ray flux from HAWC and KASCADE are already close to this line, and LHAASO should soon provide better sensitivity.

As a next step, more detailed theoretical efforts will be needed to understand the Milky Way's high-latitude gamma-ray emission. This would require multi-messenger work, including further understanding of diffuse neutrino emission.


\section{Auxiliary Multi-Messenger Strategies}
\label{sec:future}

In this section, we briefly discuss multi-messenger strategies to complement direct observations of CR electrons.  As shown in Fig.~\ref{fig:flow}, if we observe an apparent end to the CR electron spectrum, there are several possible causes.  And even if the CR electron spectrum is detected to very high energies, we still will not know the details of the various factors that lead to the observed flux.  Therefore, other types of data are needed to comprehensively disentangle the details of acceleration, escape, propagation, and detection.

Figure~\ref{fig:future-MM} outlines some likely fruitful approaches.


\subsection{Sources}

The maximum energies of electrons accelerated in individual sources is most directly probed by observations of the emission in X-rays (through synchrotron radiation) and gamma rays (through inverse-Compton scattering). Existing data indicate that pulsars are efficient accelerators of electrons, reaching energies of at least 1~PeV for the Crab~\cite{2020MNRAS.491.3217K, 2021ApJ...922..221L}. LHAASO sources seem to contain many other leptonic PeV accelerators~\cite{2021Natur.594...33C, 2023arXiv230517030C, 2023PhRvD.107d3002S}; understanding them is key to probing the maximum electron energies accelerated by sources. Future gamma-ray observations with LHAASO and SWGO~\cite{2019arXiv190208429A} will be especially important due to their sensitivity at very high energies. 
 
It is also important to test particle escape from sources.  Future gamma-ray observations with LHAASO will be especially important for probing TeV halos, which have low intensities due to their large angular extent.  As a complement, future gamma-ray observations with CTA~\cite{2019scta.book.....C} will allow studying the morphology of emitting regions, probing particle escape processes in detail.  


\subsection{Propagation}

CR hadron data offer great insights about CR propagation.  We expect the propagation of CR hadrons and electrons in the galactic magnetic field to be virtually identical, except for the rate of cooling, which is much higher for CR electrons.  The CR hadron data have the advantage that the fluxes are large, even for subdominant components such as nuclei (both stable and unstable) and antiprotons.  Secondary-to-primary ratios are important for measuring the global diffusion coefficient and the column densities for interaction and escape.  Future measurements with LHAASO, which is also a cosmic-ray detector, will be important for their increased precision.  The diffuse emission of gamma rays and neutrinos due to CR hadron interactions in the interstellar medium is also important for studying propagation effects.  It may also be possible to study CR electron propagation through the diffuse synchrotron emission that they produce. 

CR anisotropies are another rich source of information and are important for testing isotropic versus anisotropic diffusion. For hadrons, the amplitude, phase, and energy dependence of the dipole anisotropy has been measured~\cite{2015ApJ...809...90B, 2016ApJ...826..220A, 2017ApJ...836..153A, 2018ApJ...865...57A, 2018ApJ...853L..29A}. The results are difficult to reconcile with the standard isotropic diffusion models; investigating this problem has provided significant insights into the nearby sources and properties of local diffusion tensor~\cite{2012PhRvL.108u1102E, 2014ApJ...785..129K, 2015PhRvL.114b1101M, 2016PhRvL.117o1103A, 2017PrPNP..94..184A, 2019JPhCS1181a2039K, 2014Sci...343..988S, 2015ApJ...809L..23S, 2020ApJ...889...97Z, 2020ApJ...903...69F, Zhao:2021yzf}, which is a key to understanding the CR electrons. Medium- and small-scale anisotropies have also been detected in hadronic CR data.  These are useful for studying the properties of turbulent magnetic fields in the local interstellar medium~\cite{2012PhRvL.109g1101G, 2016ApJ...830...19L}.


\subsection{Detection}

The greatest unknown affecting CR electron detection is the level of the astrophysical foregrounds, especially Milky Way gamma rays at high latitude.  To reduce uncertainties, a key step is to first better understand the emission from the galactic plane at low latitudes, including a better separation between source and diffuse fluxes.  Key experiments for the gamma-ray measurements include HERD~\cite{2022PhyS...97e4010K, Berti:2023lah}, CTA~\cite{2019scta.book.....C}, and LHAASO~\cite{2019ICRC...36..471W, LHAASO:2019qtb}.  We also need a consistent picture that relates the fluxes of CR, gamma rays, and neutrinos over a wide range of energies~\cite{2018JCAP...07..006E, Sudoh:2022sdk, Sudoh:2023qrz, 2023MNRAS.518.5036M}.  With this in hand, similar studies are needed for the high-latitude emission.  Key experiments for the cosmic-ray measurements include HERD, CTA, and LHAASO, and key experiments for the neutrino measurements include IceCube~\cite{IceCube:2023ame, 2023arXiv230707576A}, IceCube-Gen2~\cite{IceCube-Gen2:2020qha}, and KM3NeT~\cite{KM3Net:2016zxf}.


\section{Conclusions}
\label{sec:summary}

Questions about the origins of CRs are manifold and longstanding, primarily because of magnetic deflections during propagation obscure the directions of sources.  For Milky Way CRs, the questions associated with CR electrons are particularly challenging, in part because their fluxes are low compared to those for CR hadrons. CR electrons have higher energy loss rates, which has the benefit that only nearby sources contribute to the observed flux.  In addition, recent rapid advances in detector sensitivity mean that CR electrons can be studied with new reach and precision, both directly and with auxiliary multi-messenger data.

Anticipating significant progress within the next several years, here we focus on the high-energy end of the CR electron spectrum, a question that has received less attention than it should.  While CR electrons have only been observed up to 5~TeV (possibly 20~TeV), it has recently been shown that multiple Milky Way sources must accelerate electrons up to at least 1 PeV.  The fact that PeV CR electrons have not been observed could be due to one of four steps that we explore: acceleration, escape, propagation, and detection.  If one or more of these steps ``fails," then the CR electron spectrum can end at energies well below 1 PeV.  We show that for each of these steps, there are significant but not unbounded uncertainties.  While we must make approximations to calculate the range of possibilities for each step, an overall clear vista emerges where decisive progress is within reach.

Determining the energy corresponding to the end of the CR electron spectrum will provide new insights about many aspects of CRs.  This may allow astronomy without directionality if it can be shown that only one nearby object could reach high enough energies.  While this may be less decisive than we might like, there are presently no clearly identified sources to the CR flux at Earth.

At still higher energies, there are exciting opportunities for exotic-physics searches, as no conventional astrophysical source should be able to contribute due to the small CR electron horizon distance at very high energies.  This should allow excellent sensitivity for searches for new physics such as dark matter annihilation or decay, which would be present at all distances from Earth.  Such signals may be enhanced by nearby dark-matter clumps.  We leave detailed investigations of these points to future work. 


\section*{Acknowledgements}

We are grateful for helpful discussions with Ivan Esteban, Carmelo Evoli, Isabelle John, Matt Kistler, Tim Linden, and Hasan Y\"{u}ksel. We thank Chris Hirata for discussions on the superluminal diffusion problem, and Ivan Esteban for discussions on the dark-matter searches. 

T.S.\ was primarily supported by an Overseas Research Fellowship from the Japan Society for the Promotion of Science (JSPS), a JSPS PD Research Fellowship, and KAKENHI Grant No.\ 23KJ2005. T.S.\ was partially supported by and J.F.B.\ was supported by National Science Foundation Grant No.\ PHY-2012955.

 
\bibliography{vhe_electrons}

\end{document}